\documentclass[times,8pt]{wlscirep}

\usepackage[utf8]{inputenc}
\usepackage[T1]{fontenc}
\usepackage{bm}
\usepackage{multicol}
\usepackage{float}

\usepackage[displaymath,mathlines]{lineno}
\title{
Measured proton electromagnetic structure deviates from theoretical predictions}

\author[1]{R. Li} 
\author[1,**]{N. Sparveris}
\author[1]{H. Atac}
\author[2]{M.K. Jones}
\author[3]{M. Paolone}
\author[17]{Z. Akbar}
\author[8]{C. Ayerbe Gayoso}
\author[5]{V. Berdnikov}
\author[6,19]{D.Biswas}
\author[1,19]{M. Boer}
\author[2]{A. Camsonne}
\author[2]{J. -P. Chen}
\author[2]{M. Diefenthaler}
\author[1]{B. Duran       }
\author[7]{D. Dutta       }
\author[2]{D. Gaskell     }
\author[2]{O. Hansen      }
\author[9]{F. Hauenstein  }
\author[10]{N. Heinrich   }
\author[2]{W. Henry       }
\author[5]{T. Horn        }
\author[10]{G.M. Huber    }
\author[1]{S. Jia         }
\author[11]{S. Joosten    }
\author[7]{A. Karki       }
\author[10]{S.J.D. Kay    }
\author[10]{V. Kumar      }
\author[16]{X. Li         }
\author[8]{W.B. Li        }
\author[6]{A. H. Liyanage }
\author[2]{S. Malace      }
\author[4]{P. Markowitz   }
\author[2]{M. McCaughan   }
\author[11]{Z.-E. Meziani }
\author[12]{H. Mkrtchyan  }
\author[13]{C. Morean     }
\author[5]{M. Muhoza      }
\author[14]{A. Narayan    }
\author[15,18]{B. Pasquini   }
\author[1]{M. Rehfuss     }
\author[2]{B. Sawatzky    }
\author[2]{G.R. Smith       }
\author[16]{A. Smith      }
\author[5]{R. Trotta      }
\author[4]{C. Yero        }
\author[17]{X. Zheng      }
\author[16]{J. Zhou       }

\affil[List of affiliations]{\footnote{$^1$Temple University, Philadelphia, PA 19122, USA. $^2$Thomas Jefferson National Accelerator Facility, VA, USA. $^3$New Mexico State University, Las Cruces, NM 88003, USA. $^4$Florida International University, University Park, Florida 33199, USA. $^5$Catholic University of America , Washington, DC 20064. $^6$Hampton University , Hampton, VA 23669. $^7$Mississippi State University, Miss. State, MS 39762. $^8$The College of William and Mary, Williamsburg, VA 23185. $^9$Old Dominion University, Norfolk, VA 23529. $^{10}$University of Regina, Regina, SK S4S 0A2, Canada. $^{11}$Argonne National Laboratory, Lemont, IL 60439. $^{12}$Artem Alikhanian National Laboratory, Yerevan, Armenia. $^{13}$University of Tennessee, Knoxville, TN 37996. $^{14}$Veer Kunwar Singh University, Arrah, Bihar 802301, India. $^{15}$University of Pavia, 27100 Pavia PV, Italy. $^{16}$Duke University, Durham, NC 27708. $^{17}$University of Virginia, Charlottesville, VA, 22904. $^{18}$INFN, 27100 Pavia (PV), Italy. $^{19}$Virginia Polytechnic Institute \& State University, Blacksburg, Virginia 24061, USA. $^{**}$corresponding author: sparveri@temple.edu}}

\begin{abstract}

The visible world is founded on the proton, the only composite building block of matter that is stable in nature. Consequently, understanding the formation of matter relies on explaining the dynamics and the properties of the proton's bound state. A fundamental property of the proton involves the system's response to an external electromagnetic (EM) field. It is characterized by the EM polarizabilities~\cite{pdg} 
that describe how easily the charge and magnetization distributions inside the system are distorted by the EM field. Moreover, the generalized polarizabilities~\cite{gps} map out the resulting deformation of the densities in a proton subject to an EM field. They reveal essential information regarding the underlying system dynamics and provide a key for decoding the proton structure in terms of the theory of the strong interaction that binds its elementary quark and gluon constituents together.
Of particular interest is a puzzle in the proton's electric generalized polarizability that remains unresolved for two decades~\cite{gps}. Here we report measurements of the proton's EM generalized polarizabilities at low four-momentum transfer squared. We show evidence of an anomaly to the behaviour of the proton's electric generalized polarizability that contradicts the predictions of nuclear theory and we derive its signature in the spatial distribution of the induced polarization in the proton. The reported measurements suggest the presence of a novel, not yet understood dynamical mechanism in the proton and present significant challenges to the nuclear theory. 

\end{abstract}

\begin{document}
\flushbottom
\maketitle
\thispagestyle{empty}
\begin{multicols}{2}

\begin{figure*}[t]
\centering
\includegraphics[width=17cm]{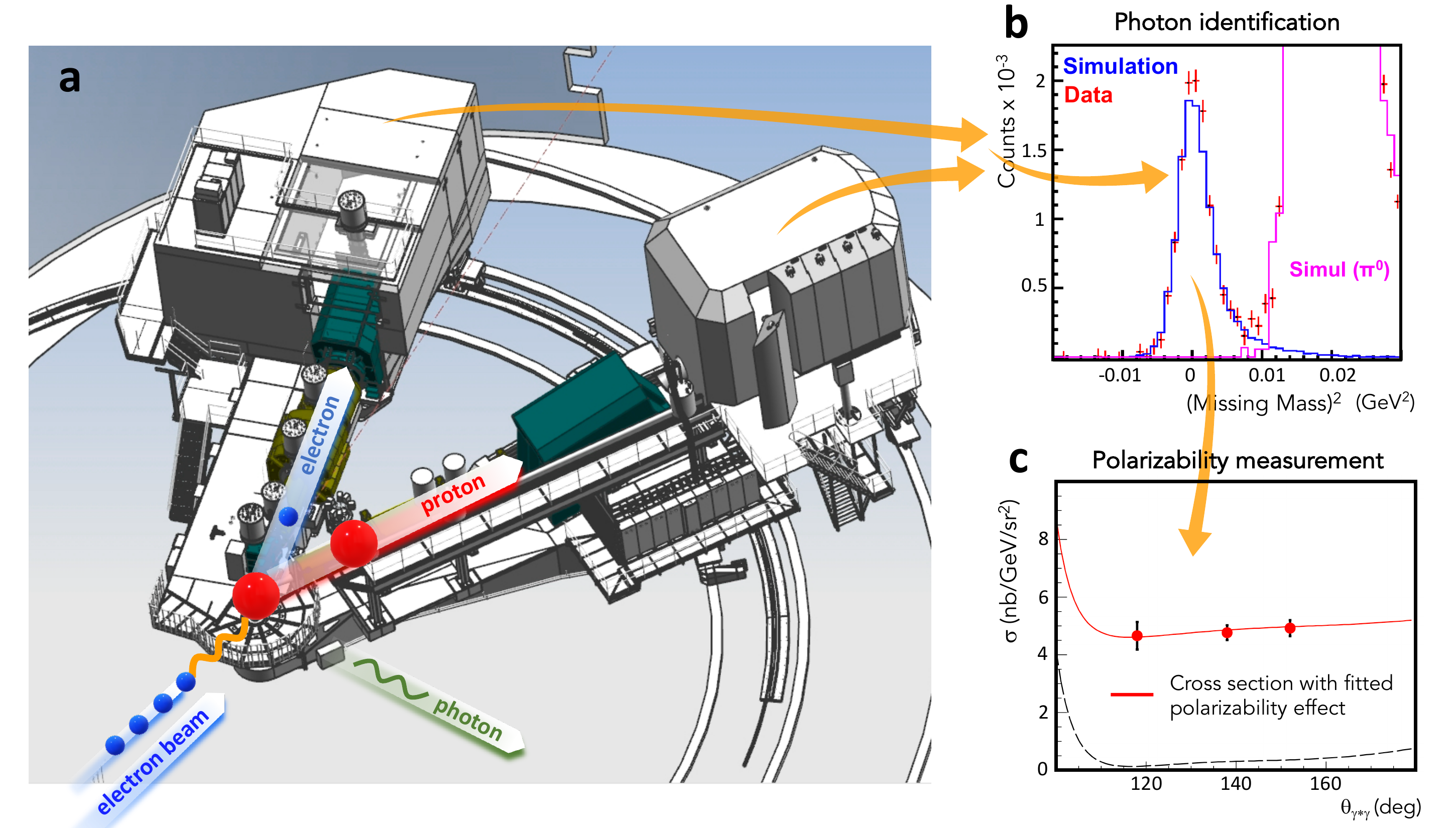}
\caption{{\bf Using virtual Compton scattering to measure the proton generalized polarizabilities} \\ {\bf a)} The experimental setup during the VCS (E12-15-001) experiment at Jefferson Lab. An electron beam impinges on a liquid hydrogen (red sphere) target. The interaction is mediated through the exchange of a virtual photon (orange wavy line). The scattered electron and recoil proton are detected with two magnetic spectrometers, in coincidence. The real photon (green wavy line) that is produced in the reaction provides the electromagnetic perturbation and allows to measure the proton polarizabilities. 
{\bf b)} The (undetected) real photon is identified through the reconstruction of the reaction's missing mass spectrum and allows the selection of the VCS events. {\bf c)} The cross section of the VCS reaction measures the proton generalized polarizabilities. The dashed line denotes the Bethe Heitler+Born contributions to the cross section. The error bars correspond to the total uncertainty, at the 1$\sigma$ or 68$\%$ confidence level.}
\label{fig-setup}
\end{figure*}

Explaining how the nucleons - protons and neutrons - emerge from the dynamics of their quark
and gluon constituents is a central goal of modern nuclear physics. The importance of the question arises from the fact that the nucleons account for 99\% of the visible matter in the universe. Moreover, the proton holds a unique role of being nature's only stable composite building block. The dynamics of quarks and gluons is governed by quantum chromodynamics (QCD), the theory of the strong interaction. The application of perturbation methods renders aspects of QCD calculable at large energies and momenta - namely at high four-momentum transfer squared ($Q^2$) - and offers a reasonable understanding of the nucleon structure at that scale. Nevertheless, in order to explain the emergence of nucleon's fundamental properties from the interactions of it's constituents, the dynamics of the system have to be understood at long distances (or low $Q^2$), where the QCD coupling constant $\alpha_s$ becomes large and the application of perturbative QCD is not possible. The challenge arises from the fact that QCD is a highly nonlinear theory, since the gluons - the carriers of the strong force - couple directly to other gluons. Here, theoretical calculations can rely on lattice QCD~\cite{lattice}, a space-time discretization of the theory based on the fundamental quark and gluon degrees of freedom, starting from the original QCD Lagrangian. An alternative path is offered by effective field theories (EFTs), such as the chiral effective field theory~\cite{cheft:1,cheft:2,cheft:3}, which employ hadronic degrees of freedom and is based on the approximate and spontaneously broken chiral symmetry of QCD. While steady progress has been made in recent years, we have yet to achieve a good understanding of how the nucleon properties emerge from the underlying dynamics of the strong interaction. In order to accomplish this, the theoretical calculations require experimental guidance and confrontation with precise measurements of the system's fundamental properties.

For a composite system, like the proton, the polarizabilities are fundamental structure constants, such as its size and shape. Listed among the system's primary properties in the Particle Data Group (PDG)~\cite{pdg}, the two scalar polarizabilities - the electric, $\alpha_E$, and the magnetic, $\beta_M$ - can be interpreted as the response of the proton's structure to the application of an external electric or magnetic field, respectively. They describe how easily the charge and magnetization distributions inside the proton are distorted by the EM field and provide the net result on the system's spatial distributions. In order to measure the polarizabilities, one must generate an electric ($\vec{E}$) and a magnetic ($\vec{H}$) field. In the case of the proton, this is provided by the photons in the Compton scattering process. The two scalar polarizabilities appear as second order terms in the expansion of the real Compton Scattering (RCS) amplitude in the energy of the photon
\begin{linenomath}\begin{equation}\label{polarizabilities} 
H^{(2)}_{eff} = -4 \pi (\frac{1}{2} \alpha_E \vec{E}^2 + \frac{1}{2} \beta_M \vec{H}^2).
\end{equation}\end{linenomath}
One can offer a simplistic description of the polarizabilities through the resulting effect of an electromagnetic perturbation applied to the nucleon constituents. An electric field moves positive and negative charges
inside the proton in opposite directions. The induced electric
dipole moment is proportional to the electric field, and the
proportionality coefficient is the electric polarizability which
quantifies the stiffness of the proton. On the other hand, a magnetic
field has a different effect on the quarks and on the pion cloud within the nucleon,  giving rise to two different contributions in the magnetic polarizability, a paramagnetic and a diamagnetic contribution, respectively.
Compared to the atomic polarizabilities, which are of the size of the atomic volume, the proton electric
polarizability $\alpha_E$ is much smaller than the volume
scale of a nucleon~\cite{pdg}. The small magnitude underlines the stiffness of the proton,
a direct consequence of the strong binding of its
constituents, and indicates the intrinsic relativistic character of the system.

The generalization~\cite{gps} of the two scalar polarizabilities in four-momentum transfer space, $\alpha_E (Q^2)$ and $\beta_M(Q^2)$, is an extension of the static electric and magnetic polarizabilities obtained in RCS.
They can be studied through measurements of the virtual Compton scattering (VCS) process~\cite{gps} $\gamma^{*}$p$\rightarrow$ p$\gamma$. The VCS is accessed experimentally through the ep$\rightarrow$ep$\gamma$ reaction. The definition of the reaction's kinematical parameters is given in the Methods section. Here, the incident real photon of the RCS process is replaced by a virtual photon. The virtuality of the incident photon $(Q^2)$ sets the scale of the observation and allows one to map out the spatial distribution of the polarization densities in the proton, while the outgoing real photon provides the EM perturbation to the system.
The meaning of the generalized polarizabilities (GPs) is analogous to that of the nucleon form factors. Their Fourier transform will map out the spatial distribution density of the polarization induced by an EM field. They probe the quark substructure of the nucleon and offer unique insight to the underlying nucleon dynamics. 
The interest on the GPs extends beyond the direct information that they provide on the dynamics of the system. They frequently enter as input parameters in various scientific problems. One such example involves the hadronic two-photon exchange corrections, which are needed for a precise extraction of the proton charge radius from muonic Hydrogen spectroscopy measurements~\cite{annrev:1}.

\begin{figure}[H]
\centering
\includegraphics[width=\linewidth]{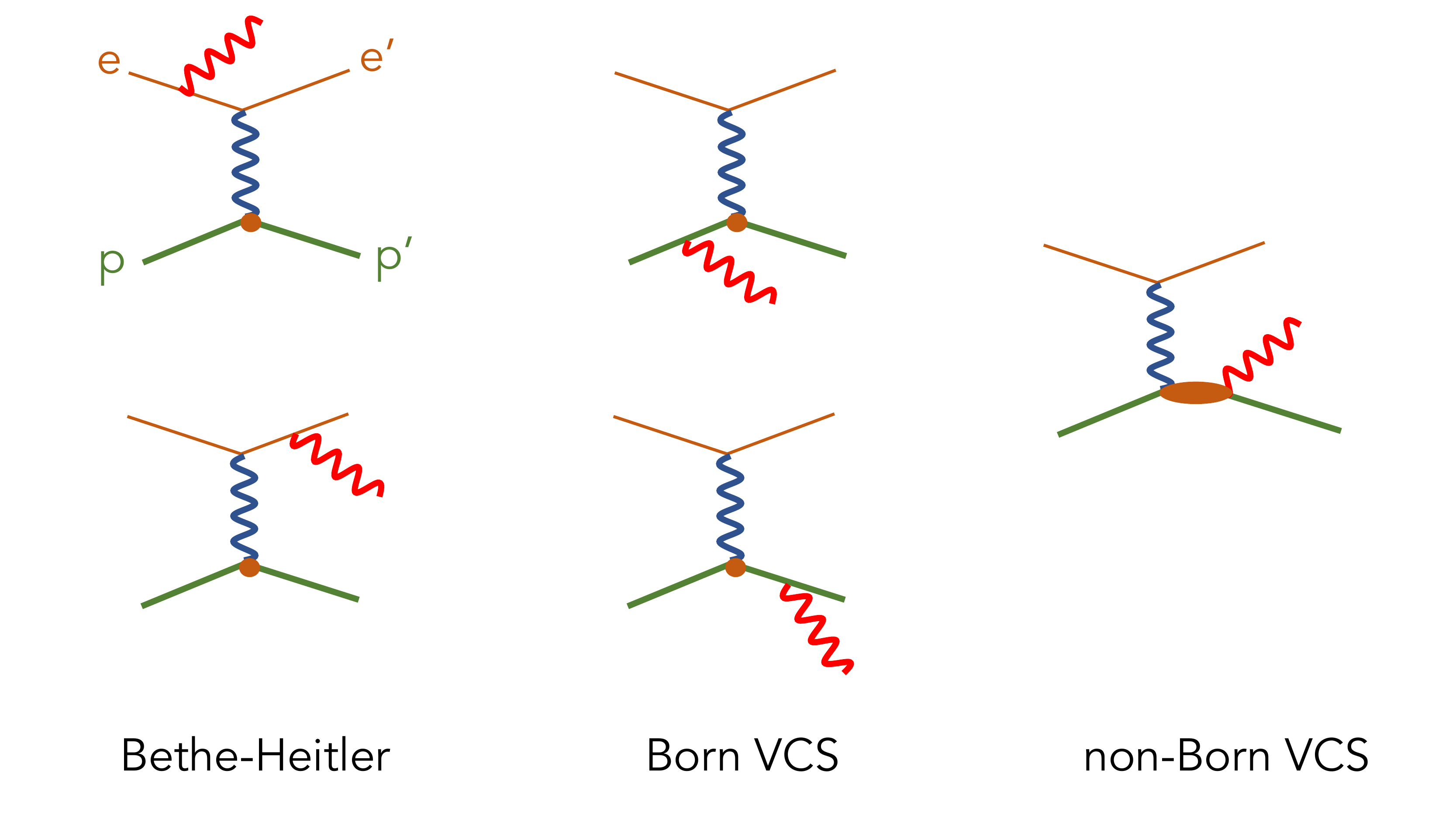}
\caption{{\bf Feynman diagrams of photon electroproduction} \\ 
The mechanisms contributing to ep$\rightarrow$ep$\gamma$. The small circles represent the interaction vertex of a virtual photon with a proton considered as a point-like particle, while the ellipse denotes the non-Born VCS amplitude.}
\label{fig-feynm}
\end{figure}

In this work, we report on measurements of the VCS reaction at the Thomas Jefferson National Accelerator Facility (Jefferson Lab). The experiment accessed the region $Q^2$=0.28~GeV$^2$ to 0.40~GeV$^2$, where the two scalar GPs are particularly sensitive to the nucleon dynamics, and aims to address a long-standing puzzle in the proton's electric GP. 
A first indication of an anomaly in this property, a local enhancement of the electric polarizability as a function of the distance-scale in the system, was reported by a measurement (later repeated by the same group) at
$Q^2$=0.33~GeV$^2$~~\cite{exp:1,exp:2} albeit with a large experimental uncertainty. Nevertheless, this anomaly has been questioned for many years. The theoretical calculations are unable to account for such a feature in the $\alpha_E(Q^2)$ and instead predict a monotonic fall-off with $Q^2$. Recent experiments have attempted to explore further the existence of such an effect with measurements that extend around the kinematical regime of interest but have not succeeded to present any supporting evidence of such a puzzling behavior in
this fundamental property~\cite{exp:7,exp:8}. This has left open a scenario that could involve issues in the experimental measurement at $Q^2$=0.33~GeV$^2$~~\cite{exp:1,exp:2} as an explanation to this problem.
In lack of an independent experimental confirmation or of further evidence, the existence of this anomaly and it's dynamical origin remains an unresolved puzzle until this day.
In this work, we capitalize on the unique capabilities of the experimental setup at Jefferson Lab along with a combination of new features in the experimental methodology to conduct measurements of the scalar GPs with unprecedented precision, targeting explicitly the kinematical
regime that is relevant to this conjectured anomaly. A first advantage of the experiment is that it exploits the sensitivity of the polarizabilities to the excited spectrum of the nucleon, that is e.g. different compared to the nucleon elastic form factors that describe only the ground state of the system. The measurements were conducted in the nucleon resonance region. 
This enables enhanced sensitivity to the polarizabilities compared to previous experiments~\cite{exp:1,exp:2,exp:7,exp:8} that measured in the region of the pion production threshold. This has been previously exhibited e.g. in~\cite{exp:3,exp:4}. 
Furthermore, in this experiment the methodology employed cross section measurements at azimuthally symmetric kinematics in the photon angle, namely for $(\phi_{\gamma^*\gamma}, \pi - \phi_{\gamma^*\gamma})$. 
The measurement of the azimuthal asymmetry in the cross section enhances even further the sensitivity in the extraction of the polarizabilities, and suppresses part of the systematic uncertainties. Moreover, the ep$\rightarrow$ep$\pi^0$ reaction was measured, simultaneously with the ep$\rightarrow$ep$\gamma$ reaction. The pion electroproduction process is well understood in this kinematic regime, and it's measurement offers a stringent, real-time normalization control to the measurement of the ep$\rightarrow$ep$\gamma$ cross section.
This offers a major enhancement to the typical normalization studies that rely on elastic scattering measurements, that we additionally also perform in this experiment.
Overall, a significant improvement was accomplished in the precision of the extracted generalized polarizabilities compared to previous measurements.

\begin{figure*}[t]
\centering
\includegraphics[width=16.5cm]{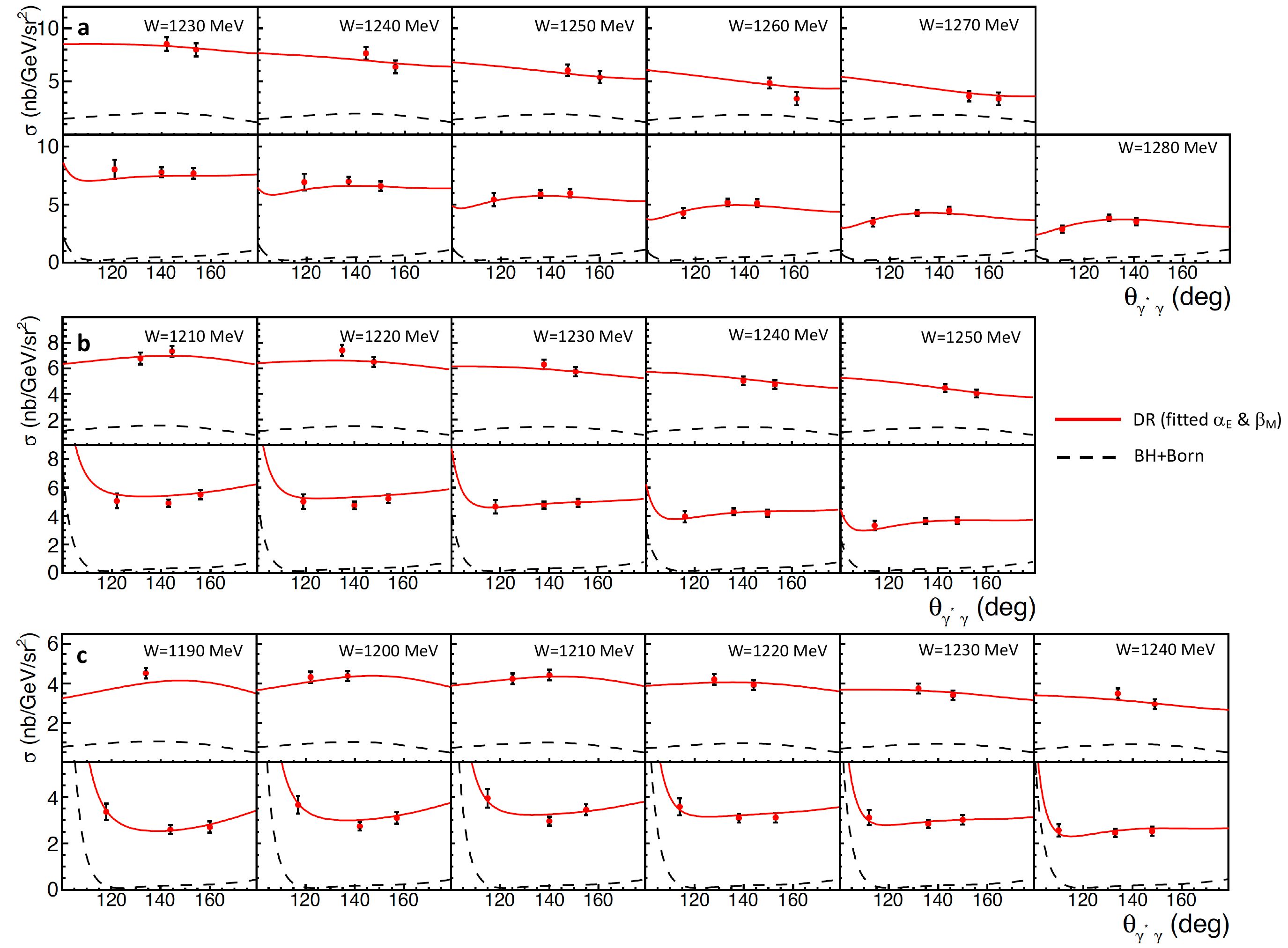}
\caption{{\bf Cross section measurements of the VCS reaction} \\ {\bf a)} Cross section measurements for in-plane kinematics at $Q^2$=0.28~GeV$^2$. Results are shown for different bins in the total c.m. energy, $W$, of the ($\gamma$p) system. {\bf b)} Measurements for in-plane kinematics at $Q^2$=0.33~GeV$^2$. {\bf c)} Measurements for in-plane kinematics at $Q^2$=0.40~GeV$^2$. Top and bottom panels correspond to $\phi_{\gamma^*\gamma}=180^\circ$ and $\phi_{\gamma^*\gamma}=0$, respectively. The solid curve shows the Dispersion Relation (DR) fit for the two scalar generalized polarizabilities. The dashed curve shows the Bethe-Heitler plus Born-VCS (BH+Born) cross section. The error bars correspond to the total uncertainty, at the 1$\sigma$ or 68$\%$ confidence level.}
\label{fig-cross}
\end{figure*}

The data were acquired in Hall C of Jefferson Lab during the VCS (E12-15-001) experiment. Electrons with energies of 4.56~GeV at a beam current up to 20~$\mu A$ were produced by Jefferson Lab’s Continuous Electron Beam Accelerator Facility (CEBAF) and were scattered from a 10~cm long liquid-hydrogen target. The Super High Momentum Spectrometer (SHMS) and the High Momentum Spectrometer (HMS) of Hall C were used to detect in coincidence the scattered electrons and recoil protons, respectively (see Fig.~\ref{fig-setup}).
Both spectrometers are equipped with similar detector packages, including a set of scintillator planes that were used to form the trigger and to
provide time-of-flight information and a pair of drift chambers used for tracking. The coincidence time was determined as the difference in the time-of-flight between the two spectrometers, accounting for path-length variation corrections from the central trajectory and for the individual start-times. The experimental setup offered a $\sim$1~ns (FWHM) resolution in the coincidence timing spectrum. Random coincidences were subtracted using the side (accidental) bands of the coincidence time spectrum. The events of the exclusive reaction ep$\rightarrow$ep$\gamma$ (see Fig.~\ref{fig-feynm}) were identified from the missing-mass reconstruction, through a selection cut around the photon peak in the missing-mass-squared spectrum. Data were taken with an empty target in order to account for the background contributions from the target walls. Elastic scattering measurements with a proton target were performed throughout the experiment for calibration and normalization studies. The measurement of the absolute VCS cross section, $\sigma \equiv d^5\sigma / d E'_{e} d \Omega'_{e} d \Omega_{cm}$, requires the determination of the coincidence acceptance, where $d E'_{e}$, $d \Omega'_{e}$ is the differential energy and solid angle of the scattered electron in the laboratory frame and $d\Omega_{cm}$ is the differential solid angle of the photon in the center-of-mass frame. The experimental acceptance is calculated with the Hall C Monte Carlo simulation program, SIMC, which integrates the beam configuration, target geometry, spectrometer acceptances, resolution effects, energy losses and radiative corrections.
The cross section results for in-plane kinematics are presented in Fig.~\ref{fig-cross}. 
The measurements are shown for different bins in the total c.m. energy, $W$, of the ($\gamma$p) system. They span across an extended range of $\theta_{\gamma^*\gamma}$ and avoid the kinematics dominated by the Bethe-Heitler process where the polarizability effect is suppressed. The complete dataset of the measured cross sections is included in the Extended Data tables.

\begin{figure*}[t]
\centering
\includegraphics[width=16.5cm]{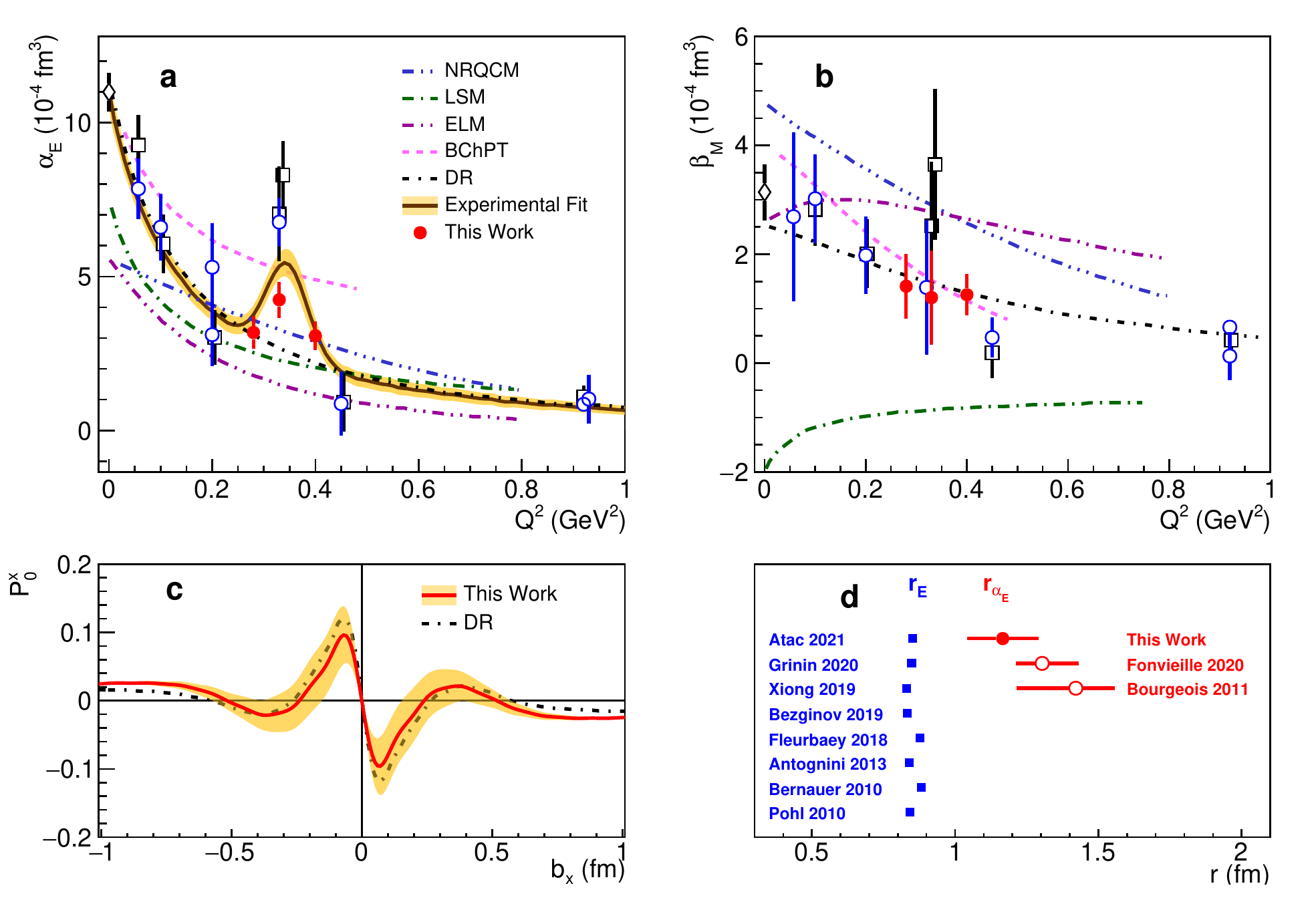}
\caption{{\bf The generalized polarizabilities of the proton} \\ {\bf a)} The electric generalized polarizability measured in this experiment (red circles). The world data~\cite{exp:1,exp:2,exp:3,exp:4,mamircs,exp:5,exp:52,exp:6,exp:7,exp:8} (open-symbols) are shown for results that involve the Dispersion-Relations (circle) and Low-Energy-Expansion analysis (box). The theoretical calculations of BChPT~\cite{ceft:1}, NRQCM~\cite{nrqm:2}, LSM~\cite{ls:1}, ELM~\cite{elm} and DR~\cite{dr:1,dr:2,dr:3} are also shown. The experimental fit that includes all the world data is also shown.
{\bf b)} The magnetic generalized polarizability. The definition of symbols and curves are the same as in (a). {\bf c)} Induced polarization in the proton when submitted to an EM field as a function of the transverse position with photon polarization along the x axis for $b_y=0$. The x-y defines the transverse plane, with the z axis being the direction of the fast moving protons. {\bf d)} The proton electric polarizability radius $r_{\alpha_E} \equiv \sqrt{\langle r^2_{\alpha_E} \rangle}$  derived from this work (red point). The measurements of the proton charge radius $r_E$ ~\cite{pohl,antognini,xiong,fleurbaey,bezginov,bernauer,atac,grinin} (blue points) are shown for comparison.
The error bars and the uncertainty bands correspond to the total (statistical+systematic) uncertainty, at the 1$\sigma$ or 68$\%$ confidence level.}
\label{fig-polar}
\end{figure*}

The cross section of the $ep\rightarrow ep\gamma$ process observes the photon that is emitted by either the lepton, known as the Bethe-Heitler (BH) process, or by the proton, the fully virtual Compton scattering (FVCS) process, as shown in Fig.~\ref{fig-feynm}. The FVCS amplitude can in-turn be decomposed into a Born contribution, with the intermediate state being the nucleon, and a non-Born contribution, that carries the physics of interest and allows for excited intermediate states of the nucleon, that is parametrized by the GPs. The BH and the Born-VCS contributions are well known, calculable in terms of the proton electromagnetic form factors that are precisely measured from elastic electron scattering. We extract the GPs from the measured cross sections through a fit that employs the dispersion relation (DR) model~\cite{dr:1,dr:2,dr:3} for VCS. In the DR formalism, the two scalar GPs enter unconstrained and can be adjusted as free parameters, while the proton electromagnetic form factors are introduced as an input. The experimental cross sections are compared to the DR model predictions for all possible values for the two GPs, and the $\alpha_E (Q^2)$ and $\beta_M (Q^2)$ are fitted by a $\chi^2$ minimization. The extracted electric and magnetic GPs are shown in Fig.~\ref{fig-polar}. We observe evidence of a local enhancement of $\alpha_E (Q^2)$ in the measured region, at the same $Q^2$ as previously reported in~\cite{exp:1,exp:2}, but we find a smaller magnitude and measure it with a significantly improved precision. The world-data at this $Q^2$ reconcile at the $\sim 2\sigma$ level. The $Q^2$-dependence of the electric GP is explored using two methods, one that employs traditional fits to the data using predefined functional forms and another one that is based on a data-driven technique that assumes no direct underlying functional form (see Methods for details). In both cases, as shown in the Extended Data Fig.~2, we find a $Q^2$-dependence for $\alpha_E(Q^2)$ that is statistically consistent with the presence of a structure in the measured region. The empirical fit to the world data is shown in Fig.~\ref{fig-polar}(a). This observation is in sharp contrast with the current theoretical understanding that suggests an $\alpha_E (Q^2)$ that decreases monotonically as the distance scale becomes smaller, namely with increasing $Q^2$. The theory predictions cover a wide range of approaches such as chiral effective ﬁeld theories~\cite{ceft:1,ceft:2,ceft:3,ceft:4,ceft:5,ceft:6}, the linear $\sigma$-model~\cite{ls:1,ls:2}, the Effective Lagrangian Model~\cite{elm}, relativistic~\cite{rqm:1} and nonrelativistic~\cite{nrqm:1,nrqm:2} constituent quark models.

The $\beta_M(Q^2)$ is expected to have a smaller magnitude relative to $\alpha_E (Q^2)$. This can be explained by the competing paramagnetic and diamagnetic
contributions in the proton, which largely cancel. In some theoretical calculations, $\beta_M(Q^2)$ is predicted to go through a maximum before decreasing. This last
feature is typically explained by the dominance of diamagnetism due to
the pion cloud at long distance (or small $Q^2$) and the dominance
of paramagnetism due to a quark core at short distance scales. 
For $\beta_M (Q^2)$, we find a smooth $Q^2$-dependence and the near-cancellation of the paramagnetic and the diamagnetic contributions in the proton at $\sim Q^2$=0.4~GeV$^2$. 
The theoretical predictions for the two generalized polarizabilities vary noticably in magnitude.
The reported measurements impose strict constraints and provide new input to the theory. 
The highlighted observation involves the puzzling $Q^2$-dependence of $\alpha_E (Q^2)$, as reflected by the fits to the world data. It manifests as a local deviation from the single-dipole behavior that describes the rest of the world data, as discussed in~\cite{exp:7,exp:8}. It contradicts the theoretical calculations, that unanimously predict a smooth fall-off as a function of $Q^2$ (see Methods section for details). 
The data add supporting evidence for the presence of a dynamical mechanism in the proton that is currently not accounted for in the theory. This would involve a dynamical element that can explain how a local enhancement of the system's electric polarizability can emerge as the distance scale becomes smaller, namely where the quark degrees of freedom acquire an increasingly prominent role in the dynamics of the system.

From the measurements of the generalized polarizabilities, we derive the spatial deformation of the quark distributions in the proton subject to the influence of an external electromagnetic field~\cite{dens:1} (see Methods for details). This follows effectively an extension of the formalism to extract the light-front quark charge densities~\cite{neutron} from the proton form factor data.
First, we derive an accurate $Q^2$-parametrization of the polarizabilities from a fit to the experimental data. From that, we extract the induced polarization in the proton, $P_0$, following Ref.~\cite{dens:1}. As shown in Fig.~\ref{fig-polar}(c), we observe that the enhancement of $\alpha_E (Q^2)$ is translated to a distinct structure in the spatial distribution of the induced polarization in the proton. The distribution follows a change of sign around $\sim$0.25~fm and exhibits a secondary maximum in the amplitude around $\sim$0.35~fm. A primary measure that quantifies the extension of a spatial distribution is the mean square radius. The mean square electric polarizability radius of the proton $\langle r^2_{\alpha_E} \rangle$ is related to the slope of the electric GP at $Q^2=0$ by
\begin{linenomath}\begin{equation}\label{radius} 
\langle r^2_{\alpha_E} \rangle = \frac{-6}{\alpha_E(0)} \cdot \frac{d}{dQ^2} \alpha_E(Q^2) {\bigg \vert}_{Q^2=0}.
\end{equation}\end{linenomath}
We determine the slope of $\alpha_E (Q^2)$ at $Q^2=0$ from fits to the world-data, using a group of functional forms that can provide a reliable fit (see Methods for details). For $\alpha_E(0)$, we adopt the most recent measurement from Ref.~\cite{mamircs}.  For the mean square electric polarizability radius we find  $\langle r^2_{\alpha_E}\rangle=1.36 \pm 0.29~fm^2$. This value is considerably larger compared to the mean square charge radius of the proton, $\langle r^2_{E} \rangle \sim0.7~fm^2$~~\cite{pdg} (see Fig.~\ref{fig-polar}(d)). The dominant contribution to this effect is expected to arise from the deformation of the mesonic cloud in the proton under the influence of an external EM field.
We derive the mean square magnetic polarizability radius from the magnetic polarizability measurements, following a procedure that is equivalent to the extraction of the mean square electric polarizability radius (see Methods for details) and we find that $\langle r^2_{\beta_M}\rangle=0.63 \pm 0.31~fm^2$.

In conclusion, we have studied the proton's response to an external electromagnetic field and its dependence on the distance scale within the system. We show evidence of a local enhancement in the proton's electric generalized polarizability that the nuclear theory cannot explain. We provide a definitive answer to the existence of an anomaly in this fundamental property and we have measured with high precision the magnitude and the dynamical signature of this effect. The reported data suggest the presence of a dynamical mechanism in the system that is currently not accounted for in the theory. 
They pose a challenge to the chiral effective field theory, the prevalent effective theory for the strong interaction, and they serve as high-precision benchmark data for the upcoming lattice quantum chromodynamics calculations. The measurements of the proton's electromagnetic generalized polarizabilities complement the counterpart of the spin-dependent generalized polarizabilities of the nucleon~\cite{natphys1,natphys2,natphys3}. Together, the two components of the generalized polarizabilities provide a puzzling picture of the nucleon's dynamics that emerge at long-distance scales.
Proton has the unique role of being nature's only stable composite building block. Consequently, the observed anomaly in a fundamental system property comes with a unique scientific interest. It calls for further measurements so that the underlying dynamics can be mapped with precision and highlights the need for an improved theory so that a fundamental property of the proton can be reliably described.



\newpage

\section*{Methods}

{\bf Experimental setup and measurements.} For the measurement of the ep$\rightarrow$ep$\gamma$ reaction, electrons with energies of 4.56~GeV at a beam current up to 20~$\mu A$ were produced by Jefferson Lab’s Continuous Electron Beam Accelerator Facility (CEBAF). The electrons were scattered from a 10~cm long liquid-hydrogen target at a temperature of 19 K. The thickness of the aluminum target cell at the entrance and exit is 0.150 (11) mm and 0.191 (19) mm, respectively. 
For every kinematical setting, data were taken with a target made of two aluminum foils located at the positions of the cryotarget entrance 
and exit windows, each having a thickness of 0.6463(10) mm, in order to subtract the background contributions emerging from the target walls by scaling the thicknesses of the two targets.
The scattered electron and recoil proton of the reaction are detected with two magnetic spectrometers, in coincidence. 
The outgoing photon of the VCS process is identified through the reconstructed missing mass spectrum. 
The polar angle $\theta_{\gamma^*\gamma}$ of the VCS reaction is defined as the center-of-mass (c.m.) polar angle of the real photon with respect to the momentum transfer direction.
The azimuthal angle of the reaction $\phi_{\gamma^*\gamma}$ defines the angle between the plane of the two (incoming and scattered) electrons and the photon-proton plane. 
The four-momentum of the outgoing photon, denoted by $\mathbf q'$, is reconstructed as $\mathbf{q'=k+p-k'-p'}$, where $\mathbf{k}$ and $\mathbf{p}$ are the four-momenta of the incoming electron and the target proton, while $\mathbf{k'}$ and $\mathbf{p'}$ are the four-momenta of the final electron and proton, respectively. The four-momentum of the virtual photon is $\mathbf{q=k-k'}$, with $Q^2 \equiv \mathbf{-q^2}$.

The beam properties were monitored throughout the experiment with the Hall C beam diagnostic elements. The beam position monitors (BPMs), that consist of a 4-wire antenna array of open ended thin wire striplines tuned to the RF frequency of the beam, were used to determine the position and the direction of the beam on the experimental target point. The beam current monitors (BCMs), a set of resonant-cavity based beam-current monitors and a parametric current transformer monitor, were used for the continuous non-intercepting beam current measurements. The beam size was measured by using harp scanners, which moved a thin wire through the beam. The beam was rastered over a 2$\times$2 mm$^2$ area to avoid overheating the target. The beam energy was determined with an uncertainty of 0.06\% by measuring the bend angle of the beam, on its way into Hall C, as it traversed the Hall C arc dipole magnets. The total accumulated beam charge was determined with 0.5\% uncertainty. The liquid-hydrogen target density receives contributions from both the target temperature and target boiling effects. The density of the liquid hydrogen target has a nearly linear dependence on the temperature. The temperature is 19 K $\pm$ 0.03 K (intrinsic electronics noise) $\pm$0.05 K (systematic uncertainty), resulting to a target density of 0.0725$\pm$0.0003 $g/cm^3$. For the target boiling effects, a correction was applied to account for the change in the target density caused by beam heating, resulting to a density fluctuation of 0.7\% at the maximum current of 20 $\mu A$ used in the experiment. The target length is measured to be 100 $\pm$ 0.26~mm thus resulting to a 0.26$\%$ uncertainty to the cross section measurement.

Two magnetic spectrometers, the Super High Momentum Spectrometer (SHMS) and the High Momentum Spectrometer (HMS) were used to detect, in coincidence, the scattered electrons and recoil protons, respectively. Both spectrometers involve a series of superconducting magnets, including quadrupoles and dipoles, followed by a set of particle detectors. 
The dipole magnets deﬂect charged particles vertically as they enter the detector huts, while the quadrupole magnets optimize the ﬂux of the charged particles entering the dipole magnet and focus the orbits of the charged particles into the detector huts. The two spectrometers are equipped with similar detector packages, with some differentiation due to the different momentum ranges of the spectrometers. The SHMS is also equipped with a Pb-glass calorimeter~\cite{pbglass} that can serve as a particle identification detector.  A pair of drift chambers, each with 6 wire planes separated by about a meter, was used to provide the tracking of the detected particles. The uncertainty in the determination of the tracking efficiency was 0.5$\%$ and 1$\%$ for the SHMS and the HMS, respectively.
A set of hodoscope planes was used to form the trigger and to
provide time-of-flight information. The time-of-flight in the HMS spectrometer was used for the proton identification, providing a $>$~20~ns separation from kaons and pions. 
The trigger efficiency of both spectrometer arms is at the 99.9\% level and comes with a $\pm0.1\%$ uncertainty.
For the correction due to the proton absorption in the spectrometer, elastic hydrogen data was taken to determine the fractional loss of protons due to inelastic collisions with material as the proton travelled from the target to the focal plane hodoscope. The fractional loss was determined with an uncertainty of 0.20\%. This correction was applied to the data and the error was included in the systematic uncertainty of the measurement.
The particle tracks are traced, through the spectrometer optics, to the target to provide the particle momentum, scattering angle and target position information. Both spectrometers offer a better than $0.1\%$ momentum resolution and an angular resolution of $\sim$ 1 mrad. The determination of the scattering angle for the SHMS and the HMS spectrometers comes with a 0.5~mrad uncertainty that is determined from constraints on the elastic kinematic reconstruction.

The coincidence time was determined as the difference in the time-of-flight between the two spectrometers, accounting for path-length variation corrections from the central trajectory and for the individual start-times. The experimental setup offered a better than 1~ns (FWHM) resolution in the coincidence timing spectrum that was measured within an 80~ns timing window. Random coincidences were subtracted using the side (accidental) bands of the coincidence time spectrum. The live-time correction, that accounts for the electronics and computer dead-time, came with an uncertainty that ranged between 0.3$\%$ and 0.6$\%$ for the different kinematical settings of the experiment. To estimate the systematic error on this correction, we used the standard deviation of the Gaussian fit to the histogram of the deadtime of the runs used in each kinematic setting. A run was normally about half an hour of beam time and the number of runs per kinematic setting ranged from about 50 to 110.

The events of the exclusive reaction ep$\rightarrow$ep$\gamma$ were identified from the missing-mass reconstruction, through a selection cut around the photon peak in the missing-mass-squared spectrum. The contamination from the missing mass tail of the pion electroproduction events was studied via two methods. The contributions were evaluated by pion electroproduction simulation studies that employ the well known cross section of the reaction and offer an accurate description of the missing mass tail. In a second method, the pion contamination was determined through a phenomenological parametrization of the missing mass spectrum. The two methods exhibit an agreement in the extracted cross section at the percent level. A 1$\%$ uncertainty was assigned to this correction.

Elastic scattering measurements with a proton target were performed at different stages of the experiment, for calibration and normalization studies. A real-time normalization cross check during the measurement of the VCS cross section was also performed from the simultaneous measurement of the ep$\rightarrow$ep$\pi^0$ reaction. In both the elastic and pion electroproduction measurements, we found an excellent agreement to these well known cross sections, and confirmed that the spectrometer acceptance is accurately described by the simulation of the experiment.

The true momentum settings of the two spectrometers were determined based on a cross-calibration method that utilizes the pair of the azimuthal asymmetry measurements. Here, the momentum and position of the electron spectrometer remain the same between the two kinematical settings. The momentum setting for the proton spectrometer also remains constant, while the proton spectrometer is re-positioned symmetrically with respect to the momentum transfer direction. Since the two kinematical settings involve identical momentum settings for each of the two spectrometers, the determination of their true momentum settings comes from a unique solution for both kinematics, that simultaneously calibrates the reconstructed VCS and pion electroproduction missing mass peaks to the true physical mass values for the photon and the pion, respectively. The correction between the set and the true values in the central momentum of the two spectrometers was smaller than 0.1$\%$.

{\bf Cross sections and generalized polarizabilities.} The measurement of the absolute VCS cross section, $\sigma \equiv d^5\sigma / d E'_{e} d \Omega'_{e} d \Omega_{cm}$, requires the determination of the coincidence acceptance, where $d E'_{e}$, $d \Omega'_{e}$ is the differential energy and solid angle of the scattered electron in the laboratory frame and $d\Omega_{cm}$ is the differential solid angle of the proton in the center-of-mass frame.
The determination of the coincidence acceptance is calculated by using the Monte Carlo simulation program, SIMC. The simulation integrates a realistic description of the beam configuration, target geometry, spectrometer acceptances, resolution effects, energy losses and the radiative corrections as described in~\cite{radiative:1}. The measured cross section is derived as:
\begin{linenomath}
\begin{equation}
\sigma  = \frac{N }{\epsilon_{LT} \cdot \epsilon_{trk} \cdot \epsilon_{trig} \cdot \epsilon_{p} \cdot L \cdot \Delta \Omega^{5} } f_{rad} \frac{\sigma_{P}^{sim}}{\sigma_{avg}^{sim}}
\label{eq:xs}
\end{equation}
\end{linenomath}
The parameters of Eq.~\ref{eq:xs} are as follows: $N$ is the number of the measured events within the acceptance cuts after the subtraction of the contributions arising from the target walls and from the accidental background, $\Delta \Omega^{5}$ is the five-fold coincidence solid angle that is determined from the simulation of the experiment,  $\epsilon_{LT}$ denotes the efficiency that is associated with the computer and electronics live-time, $\epsilon_{trk}$ and $\epsilon_{trig}$ are the combined tracking and trigger efficiencies for the two spectrometers, respectively and $\epsilon_{p}$ is the efficiency that corrects for the proton absorption in the spectrometer. The $f_{rad}$ denotes the radiative corrections and the luminosity $L=\frac{\rho * l*N_A}{A} * \frac{Q}{e}$, where $\rho$ is the target density in $g/cm^3$, $l$ is the target length in $cm$, $N_A$ is Avogadro's number, A is the mass number, Q is the accumulated charge of the measurement and $e$ is the electron charge.
The $\sigma_{P}^{sim}$ denotes the point cross section of the simulation at the central kinematics of each bin, while $\sigma_{avg}^{sim}$ is the average cross section of the simulation within the analysis bin. 
The term $\sigma_{P}^{sim}/\sigma_{avg}^{sim}$ effectively provides the bin-centering correction for the extraction of the point cross section from the finite phase space of the analysis bin.  
A first-layer global phase-space cut in the data analysis selects the central half of the coincidence acceptance, so that any potential influence from acceptance edge-effects is eliminated. 
The bin width size ($Q^2, W, \theta_{\gamma^*\gamma}, \phi_{\gamma^*\gamma}$) is varied in the analysis so as to validate the stability of the results as a function of the bin-size selection, and to confirm the good understanding of the coincidence phase space in the experiment simulation. The generated events in the simulation are weighted with a cross section using the DR model for VCS~\cite{dr:1,dr:2,dr:3}. In the DR formalism, the two scalar GPs come as an unconstrained part and can be adjusted as free parameters, while the proton electromagnetic form factors are introduced as an input. For the non-Born VCS part, a realistic initial parametrization is applied based on the current knowledge of the GPs. We extract the GPs from the measured cross sections through a fit that employs the dispersion relations model. The experimental cross sections are compared to the DR model predictions for all possible values for the two GPs, and the $\alpha_E (Q^2)$ and $\beta_M (Q^2)$ are fitted by a $\chi^2$ minimization. 
Resolution and experimental parameters are studied by varying them in the analysis within their quoted precision, and their effect on the measured cross sections and to the extracted GPs is quantified as a systematic uncertainty. 
Other sources of systematic uncertainties involve the radiative corrections~\cite{radiative:1} that introduce a $1.5\%$ uncertainty to the measured cross section, and the uncertainty in the determination of the coincidence solid angle that is better than $1.5\%$. The bin-centering correction was studied by varying the cross section model in the simulation and was found to be very small.
The cross section results are reported in the Extended Data Tables 1, 2 and 3. The extracted generalized polarizabilities are reported in the Extended Data Table 4.

{\bf $Q^2$-dependence of the electric GP.} 

The theoretical models that include a physical mechanism for the polarizabilities give a poor fit to the data with a a reduced-$\chi^2$ of $\chi_{\nu}^2$=7.69 (BChPT), 14.18 (NRQCM), 13.09 (LSM) and 24.06 (ELM). The DR prediction that involves an empirical parametrization for the polarizability exhibits also a poor fit with a $\chi_{\nu}^2$=5.97.

We explore the $Q^2$-dependence of the electric GP following two methods. In the first method, we explore $Q^2$-parametrizations that will offer a good description of the experimental data. We work on the basis of the two parametrizations that have been considered in the past in Ref.~\cite{dens:1}. The first function involves the typical dipole parametrization. That is a natural functional form that can effectively describe similar physics quantities, such as the nucleon elastic form factors and the magnetic GP, and can satisfy the monotonic $Q^2$-dependence that is predicted by the current theoretical models for the electric GP. We find that such a functional form does not provide a good fit to the world data and results to a reduced-$\chi^2$ of $\chi_{\nu}^2$=3.7. As seen in Extended Data Fig.~2a, the fit systematically overestimates the results from the most recent MAMI experiment (MAMI-VI)~\cite{exp:7,exp:8} that involves the most refined measurements using this experimental setup. It also systematically underestimates the two MAMI experiments (MAMI-I~\cite{exp:1} and MAMI-IV)~\cite{exp:2} at $Q^2$=0.33~GeV$^2$ and runs grossly through the new measurements that we report in this work. In order to seek a successful functional form, we follow the recipe suggested in Ref.~\cite{dens:1} and we add an additional structure that is parametrized through a Gaussian term. This empirical parametrization, as shown in Extended Data Fig.~2a, offers a fit with a $\chi_{\nu}^2$=1.9. It offers a reasonable description of the world data and is compared to the theoretical predictions in Fig.~\ref{fig-polar}(a) (denoted as Experimental Fit).
An additional method is considered using a Gaussian process regression (GPR) technique ~\cite{gauss:1} which assumes no direct underlying functional form (i.e. polynomial, exponential, Gaussian, or any combination thereof) and provides the best linear unbiased prediction for a governing distribution based on the available data.  The GPR prediction is shown in the Extended Data Fig.~2b.  Being data driven, the resulting GPR technique cannot provide a very precise prediction in the large $Q^2$ region ($Q^2 > 0.5 GeV^2$) where data is sparse.  In the lower $Q^2$ region, the GPR predicts a distribution with similar confidence as the dipole+Gaussian fit shown in the Extended Data Fig.~2a.  

{\bf Induced polarization.} 
We derive the transverse position space dependence of the induced polarization in the proton following~\cite{dens:1}:
\begin{equation}
\vec P_0(\vec b) = \hat b \, \int_0^\infty \frac{d Q}{(2 \pi)} \, Q  \, J_1(b \, Q) \,  A(Q^2),
\label{eq:indpol}
\end{equation}
where $\vec b$ is the transverse position, $b = | \vec b|$, $\hat b = \vec b / b$ and $J_1$ the 1st order Bessel function. The $A$ is a function of the GPs~:
\begin{eqnarray}
&& \hspace{-0.3cm} A = - (2 M) \, \sqrt{\tau} \,
\sqrt{\frac{3}{2}} \sqrt{\frac{1 + 2 \tau}{1 + \tau}} \nonumber \\
&& \hspace{-0.3cm} \times  \left\{ -  P^{(L1, L1)0} + \frac{1}{2} P^{(M1, M1)0} - \sqrt{\frac{3}{2}} P^{(L1, L1)1}   \right. \nonumber \\
&&\left. \hspace{-0.3cm}
- \sqrt{\frac{3}{2}} (1 + \tau) \left[ P^{(M1, M1)1} + \sqrt{2} \, (2 M \tau) P^{(L1, M2)1} \right]  
\right\} . \quad  
\label{eq:alpha}
\end{eqnarray}
The GPs are expressed in the multipole notation $P^{(\rho' \, l', \rho \,l)S}$~~\cite{multipole:1}, where 
$\rho$ ($\rho'$) refers to the Coulomb/electric ($L$), or magnetic ($M$) nature of the initial (final) photon, $l$ ($l' = 1$) is the angular momentum of the initial (final) photon, and $S$ differentiates between the  spin-flip ($S=1$) and non spin-flip ($S=0$) transition at the nucleon side. The $\tau \equiv Q^2 / (4 M^2)$, with $M$ the nucleon mass. The two scalar GPs are defined  as:
\begin{equation}
\alpha_E (Q^2) =   - {e^2 \over 4 \pi}  \sqrt{3 \over 2}~P^{(L1,L1)0} (Q^2) \\
\label{eq:conv1}
\end{equation}
\begin{equation}
\beta_M (Q^2)  =  - {e^2 \over 4 \pi}  \sqrt{3 \over 8}~P^{(M1,M1)0} (Q^2)
\label{eq:conv2}
\end{equation}
with  $e^2 / 4 \pi = 1/\alpha_{QED} =  1/137$. In calculating the $A$~function, the spin GPs are fixed by the dispersion relations~\cite{dr:2,dr:3}. For the asymptotic part of $\alpha_E(Q^2)$ we use the parametrization that we derive from the {\it Experimental Fit} to the world data:
\begin{equation}
\alpha_{E}(Q^{2}) = p0*e^{-0.5*(\frac{Q^{2}-p1}{p2})^{2}} +\frac{1}{(p3+Q^{2}/p4)^{2}}~(fm^3) \\
\label{eq:param1}
\end{equation}
with $p0=(30.4\pm6.1)10^{-5}$, $p1=0.345\pm0.008$, $p2=0.040\pm0.003$, $p3=34.217\pm1.136$ and $p4=0.014\pm0.002$.
For $\beta_M(Q^2)$ we find that the world data are described accurately by the DR model~\cite{dr:2,dr:3} that adopts a single dipole
behavior for the unconstrained part of the scalar GPs with a mass scale parameter of $\Lambda_\beta=0.5~GeV$, and we adopt this parametrization.

{\bf Electric and magnetic polarizability radius.}
The electric polarizability radius $\langle r^2_{\alpha_E} \rangle$ is extracted from Eq.~\ref{radius}. For $\alpha_E(0)$ we adopt the most recent measurement from~\cite{mamircs}.  In order to determine the slope of the electric GP at $Q^2=0$ we explore a viariety of functional forms, namely combinations of polynomial, dipole, gaussian and exponential functions. We determine those functional forms that can provide a good fit to the data and a meaningful extraction of the slope in terms of its uncertainty. The fits are explored in two groups: one over the full $Q^2$ range, and a second one within a limited range at low-$Q^2$ that does not include the $\alpha_E$ anomaly, namely for $Q^2$=[0,0.28]~GeV$^2$. For the experiments where the polarizabilities have been derived by both the Dispersion Relations and the Low Energy Expansion analysis, the variance of the two results is treated as a model uncertainty for each data point. The results of the individual fits are shown in the Extended Data Fig.~3. For each group, the final value for $\langle r^2_{\alpha_E} \rangle$ is determined from the weighted average of the results of the individual fits. The uncertainty of $\langle r^2_{\alpha_E} \rangle$ receives contributions from the uncertainty of the weighted average and from a second term that is quantified from the weighted variance of the individual fit results and effectively reflects the model dependence on the choice of the fitted parametrization. This is similar to what has been followed in the past for the extraction of the proton charge radius from fits that employ multiple functional forms e.g.~\cite{bernauer,atac}.  The final result is derived from the average of the two group values, with their spread accounted as a model uncertainty. The new result for $\langle r^2_{\alpha_E} \rangle$ updates the earlier extractions~\cite{exp:52,gps} of this quantity, as shown in Extended Data Fig.~4. In comparison to these results, the past derivations of $\langle r^2_{\alpha_E} \rangle$ ~\cite{exp:52,gps} have been performed considering a fit of a single function within a limited $Q^2$-range, and an older measurement for $\alpha_E(0)$.

The mean square magnetic polarizability radius is derived from the magnetic polarizability measurements following
\begin{linenomath}\begin{equation}\label{radius-magnetic} 
\langle r^2_{\beta_M} \rangle = \frac{-6}{\beta_M(0)} \cdot \frac{d}{dQ^2} \beta_M(Q^2) {\bigg \vert}_{Q^2=0}.
\end{equation}\end{linenomath}
For $\beta_M(0)$ we adopt the most recent measurement from~\cite{mamircs}.  
 In order to determine the slope of the magnetic GP at $Q^2=0$ we explore a viariety of functional forms, namely combinations of polynomial, dipole and exponential functions. We determine those functional forms that can provide a good fit to the data and a meaningful extraction of the slope in terms of its uncertainty e.g. for the dipole and (dipole$\cdot$polynomial) functions we find that they give a good fit, but the radius is derived with a very large uncertainty and does not influence the final extraction of this quantity.
 The results of the individual fits are shown in the Extended Data Fig.~5.
 The exponential fit employs only 2 free parameters and offers an uncertainty ($\langle r^2_{\beta_M}\rangle=0.41 \pm 0.10$~fm$^2$) that is significantly smaller compared to the other functional forms that involve 3 or more free parameters. The fitted exponential curve also appears to be systematically different compared to the rest of the fitted functions, as can be seen in the Extended Data Fig.~5. In order not to bias the final extraction of the radius by the small uncertainty (or equivalently, the large weight factor) of this one fit, the fitted results are divided into two groups, one for only the exponential fit and a second group for the rest of the functions. For the second group, we derive the radius based on the weighted average and the weighted variance of the individual fits and we find  $\langle r^2_{\beta_M}\rangle=0.85 \pm 0.25$~fm$^2$. We adopt the mean average of the two group values as the final result for the magnetic polarizability radius. For the uncertainty, we consider the spread of the two group values as a model uncertainty and we add it linearly to the statistical uncertainty. We find that $\langle r^2_{\beta_M}\rangle=0.63 \pm 0.31$~fm$^2$.

\section*{Data availability}
The raw data from the experiment are archived in Jefferson Laboratory’s mass storage silo and at Temple University, Department of Physics. The filtered data are archived at Temple University. The data are available from the authors upon request.

\section*{Code availability}
The data analysis uses the standard C++ ROOT framework, which was developed at CERN and is freely available at https://root.cern.ch. The simulation of the experiment was generated with the Jefferson Lab simulation code SIMC. The DR fit was done using the dispersion relations code developed by B. Pasquini~\cite{dr:1,dr:2,dr:3}.
The computer codes used for the data analysis are available upon request.

\section*{Acknowledgements}
We would like to thank M.~Vanderhaeghen as this work received great benefit from his input and suggestions. This work has been supported by the US Department of Energy Office of Science, office of Nuclear Physics under contracts no. DE-SC0016577 and DE-AC05-06OR23177.

\section*{Author contributions}
N.S. is the spokesperson of the experiment. He initiated, guided and supervised this effort. R.L worked on the data analysis and on the extraction of the cross sections and generalized polarizabilities. H.A. worked on the data analysis and on the extraction of the electric and magnetic polarizability radii. M.K.J. and M.P. co-supervised the data analysis efforts and worked on the simulation of the experiment. B.P. developed the DR code for VCS and produced the induced polarization density. The entire VCS collaboration participated in the data collection and in the on-line analysis of the experiment.

\section*{Competing interests}
The authors declare no competing interests. 

\section*{Supplementary information}
Supplementary information is available for this paper in the form of Extended Data figures and tables.

\section*{Correspondence and request of materials}
Correspondence and request of materials should be addressed to N.S.

\end{multicols}

\end{document}